\documentclass[a4paper, 12pt]{report}

\title{A Suggestion for a Teleological Interpretation of Quantum
Mechanics}
\author{Eyal Gruss}

\newcommand{\bra} [1] {\langle#1\vert}
\newcommand{\ket} [1] {\vert#1\rangle}
\newcommand{\up} {\uparrow}
\newcommand{\dn} {\downarrow}
\newcommand{\al} {\alpha}
\newcommand{\bt} {\beta}
\newcommand{\eps} {\varepsilon}
\newcommand{\sig} {\sigma}
\newcommand{\rhod} {\rho\mathrm{^{DEN}}}
\newcommand{\half} {\frac{1}{2}}
\newcommand{\hrt} {\frac{1}{\sqrt{2}}}
\newcommand{\ox} {\otimes}
\newcommand{\imp} {\longrightarrow}
\newcommand{\uk} {\ket{\up}}
\newcommand{\h} {\hbar}
\newcommand{\scal} [2] {\langle#1\vert#2\rangle}
\newcommand{\dk} {\ket{\dn}}
\newcommand{\ub} {\bra{\up}}
\newcommand{\db} {\bra{\dn}}
\newcommand{\ak} [1] {\ket{\mathrm{#1}}}
\newcommand{\ab} [1] {\bra{\mathrm{#1}}}
\newcommand{\sog} [1] {\left(#1\right)}
\newcommand{\ek} [2] {\ket{\mathrm{\eps_{#1}\sog{#2}}}}
\newcommand{\eb} [2] {\bra{\mathrm{\eps_{#1}\sog{#2}}}}
\newcommand{\tr} {\mathrm{tr}}
\newcommand{\prob} {\mathrm{prob}}
\newcommand{\abs} [1] {\vert#1\vert}
\newcommand{\si} [1] {\sin\sog{#1}}
\newcommand{\co} [1] {\cos\sog{#1}}
\newcommand{\e} [1] {\exp\sog{#1}}

\begin{document}

\begin{titlepage}
\begin{center}
The Hebrew University of Jerusalem\\ Racah Institute of Physics\\
\vfill
\begin{LARGE} \textbf{A Suggestion for a Teleological
Interpretation of Quantum Mechanics}\\ \mbox{}\\
\end{LARGE}
\begin{large}
Eyal Gruss\\ \medskip
\end{large}
Racah Institute of Physics\\ The Herbrew University\\ Giv'at-Ram,
Jerusalem 91904, Israel\\ e-mail: \verb+eyal_gruss@yahoo.com+\\
\vfill
\begin{tabular}{ll}
Instructors:&Prof. Yakir Aharonov, Tel-Aviv University\\ &Prof.
Issachar Unna, the Hebrew University of Jerusalem
\end{tabular}
\vfill
\begin{large}
A thesis submitted in partial fulfilment\\ of the requirements for
the degree of\\ Master of Science of the Hebrew University of
Jerusalem\\ \mbox{}\\ May 2000
\end{large}
\end{center}
\end{titlepage}

\begin{flushright}
\textit{For Keren,\\ who has always opposed\\ the probabilistic
method}
\end{flushright}
\vfill
\begin{quotation}
``I don't understand you,'' said Alice. ``It's dreadfully
confusing!''

``That's the effect of living backwards,'' the Queen said kindly:
``it always makes one a little giddy at first---''

``Living backwards!'' Alice repeated in great astonishment. ``I
never heard of such a thing!''

``--but there's one great advantage in it, that one's memory works
both ways.''

``I'm sure \textit{mine} only works one way,'' Alice remarked. ``I
can't remember things before they happen.''

``It's a poor sort of memory that only works backwards,'' the
Queen remarked.

\begin{flushright}
[Lewis Carroll, Through the Looking-Glass]
\end{flushright}
\end{quotation}
\vfill
\begin{abstract}
We suggest solving the measurement problem by postulating the
existence of a special future final boundary condition for the
universe. Although this is an extension of the way boundary
conditions are usually chosen (in terrestrial laboratories), it is
our only deviation from standard quantum mechanics. Using two
state vectors, or the ``two-state'', to describe completely the
state of systems of interest, we analyze ideal and ``weak''
measurements, and show the consistency of our scheme. If the final
state of a measuring device is assigned to be \emph{one} of the
possible outcomes of the measurement, an effective reduction is
observed after an ideal measurement process. For final conditions
chosen with an appropriate distribution, the predictions of
standard quantum mechanics may be reconstructed, thus eliminating
probability from the description of any single measurement. The
interpretation attained, the \emph{Teleological Interpretation},
is an ontological one; it is local and deterministic. Other
special assumptions in the choice of the final boundary condition
may explain certain unaccounted for phenomena, or even supply a
mechanism for essential free will. In this context we believe that
a new conception of time should be adopted.
\end{abstract}

\tableofcontents

\chapter{Introduction}
We start by presenting in Chapter \ref{ch:back} some of the
conceptual difficulties in the foundations of quantum mechanics.
They include the measurement problem, problems which arise in
various popular interpretations of quantum mechanics and the
problem of defining classicality. These fundamental issues are
subjects of many discussions taking different approaches, and we
wish to present the conceptual common ground necessary for the
understanding of our work, and put it in uniform terminology.

In Chapter \ref{ch:symm} we briefly review the time-symmetric
reformulation of quantum mechanics, in which the suggested
interpretation is formulated. In this formalism, two temporal
boundary conditions are assumed: an initial one and a final one.
Two wave-functions are evolved from these, respectively: the
standard one, which we call the history vector, and a backwards
evolving one, the hermitian adjoint of which we call the destiny
vector. These are combined to form the two-state, constituting the
complete description of any system. Final boundary conditions are
usually introduced by post-selection of the measured system,
taking into account only the experiments which yield a specific
outcome. We wish to generalize this formulation to be applicable
also to closed systems, namely to the universe. We show that the
density matrix of the closed universe in standard quantum
mechanics is a special case of its possible two-states in the
time-symmetric formalism. From here on we work in the frame of the
two state vector formalism and are interested only in the unitary
evolution of two-states and in the reduced two-states of their
subsystems.

In Chapter \ref{ch:tele} we suggest choosing a special final
boundary condition which solves the measurement problem. This is
done by setting the final states of measuring devices as
\emph{one} of their classically possible pointer states, according
to the measurements in which they are involved, with a probability
distribution that reconstructs the predictions of standard quantum
mechanics for large ensembles of such measurements. By this,
probability is eliminated from the description of any single
measurement, and its specific outcome may be calculated, given its
boundary conditions. We analyze ideal measurements and show how
effective reduction is attained without the need of supplemental
mechanisms, thus solving the measurement problem. Here we
introduce the process of ``two-time decoherence'', which, we show,
governs the behaviour of the measuring device. We next consider
non-ideal weak measurements, where a weak interaction between the
measuring device and the system being measured takes place, and
the outcome is the expected value of the operator dependent of the
boundary conditions. Clearly these kind of measurements may be
naturally described using the two-state formalism. We remark that
weak values, or the strange outcomes obtained from weak
observations, may be used to explain miscellaneous unaccounted for
phenomena.

In Chapter \ref{ch:prop} we discuss the validity of classical
properties such as locality, causality, realism, determinism and
free will, in the framework of our interpretation. This discussion
is mainly philosophical, but is also an indispensable part of a
complete picture of any interpretation. We argue that all
properties excluding realism are valid. We further suggest a
mechanism which allows essential free will within the framework of
the suggested interpretation, and claim that in this context a
more complex approach to the concept of time should be adopted.

Chapter \ref{ch:last} constitutes a discussion of ideas which
relate to the interpretation suggested, and a summary of our work.

\chapter{Background}
\label{ch:back}
\section{The Measurement Problem}

Quantum theory was originally intended to describe the behaviour
of microscopic particles, and indeed it predicts with great
accuracy the outcomes of experiments in this regime. Furthermore,
quantum theory has supplied explanations to phenomena previously
unaccounted for, such as the spectrum of black body radiation.
Today quantum theory is almost unanimously accepted as correct.

Nevertheless quantum theory raises considerable physical and
philosophical difficulties regarding the usage of classical
concepts such as locality, realism, determinism and so forth. The
difficulty intensifies when we attempt to apply quantum theory to
the macroscopic regime, meaning also to our measuring devices --
an action that seems legitimate, for also they are made of
microscopic particles. At first sight, it seems that in order to
build a theoretical model which may reconstruct the empirical
results, one needs to define an observer or a measuring device
external to the observed quantum system, one which does not obey
the normal evolution rules of quantum mechanics. This of course is
not desired, because then quantum theory will be incomplete, in
the sense that it will not describe all physical nature. This is
known as the measurement problem.

The measurement problem can be simply demonstrated in the
following manner. An experiment is performed on a spin-$\half$
particle in order to find its spin component along some axis. Let
the initial state of the particle be $\hrt\sog{\uk+\dk}$. The
initial state of the measuring device is $\ak{R}$ (device
``READY''). Now assume an interaction between the particle and the
measuring device takes place, such that if the device measures
$\uk$, it evolves into the state $\ak{U}$ (device measured
``UP''), and if it measures $\dk$, it evolves into the state
$\ak{D}$ (device measured ``DOWN''). The evolution predicted by
quantum mechanics is
\begin{equation}
\label{eq:problem}
\hrt\sog{\uk+\dk}\ox\ak{R}\imp\hrt\sog{\uk\ox\ak{U}+\dk\ox\ak{D}}.
\end{equation}
But we know from everyday practice (assume we are
experimentalists) that when measuring a spin-$\half$ particle as
above, we get $\uk\ox\ak{U}$ in $50\%$ of the cases and
$\dk\ox\ak{D}$ in the other $50\%$, not a superposition of the two
as in (\ref{eq:problem}).

In general, the experiment shows that after an ideal measurement,
the quantum system is found to be in one of the eigenstates of the
measured operator, correlated to the appropriate state of the
measuring device. For pre-selected ensembles this happens with a
probability equal to the absolute value of the projection of the
initial state on the specific eigenstate, squared.

In order to explain this gap between theory and observed reality,
different interpretations of quantum mechanics have been
suggested, or better phrased, different interpretations of our
observations. Their common goal is to provide the most complete
description of physical reality possible, and to settle some of
the contradictions between quantum theory and classical concepts.

\section{Interpretations of Quantum Mechanics}
\label{se:inter} Generally, interpretations of quantum mechanics
can be divided into two main categories. The first kind explains
our observations by supplementing the evolution rules of quantum
mechanics with the concept of a random ``collapse of the
wave-function''. A reduction of the quantum superposition state to
a classical state is supposed to take place at some stage of the
measurement process. These interpretations are problematic because
the collapse is non-local \cite{noncovar}, the collapse mechanism,
if at all specified, gives a nondeterministic outcome, and Occam's
razor, the principle of simplicity, is unsympathetic to the
excessive collapse rule.

A second kind of interpretation attempts to explain experimental
outcomes deterministically, without the need of a collapse. Among
these are ``hidden variables'' interpretations and ``relative
state'' interpretations. hidden variables interpretations,
following Bohm \cite{bohm}, assume the existence of inaccessible
variables with definite values, which determine the state after
the measurement. Bell \cite{bell}, in his famous inequality
theorem, and recently the GHZ \cite{ghz} \cite{mermin} argument,
show that these kind of theories are inherently non-local. Also it
is not clear how the empirical probabilities, as they arise in the
standard theory, are to be reconstructed. The different variations
of the relative state interpretation (such as ``many worlds'' or
``many-minds''), which are themselves interpretations of Everett's
original suggestion \cite{everett}, settle the measurement problem
by attributing different observations to the different states of
consciousness correlated to them. In these interpretations there
is no simple relation of the empirical probabilities to the
possible outcomes. In fact it is assumed that all possible
outcomes and observations coexist, so one cannot explain why
\emph{he} is the one observing a certain outcome. Both of these
non-collapse interpretations also contain some multiplicity of
entities (variables, worlds), needed for the description of the
system, against the spirit of Occam's razor.

It is worth commenting that we have discussed only ontological
interpretations whereas epistemological ones exist also. The
former are interpretations in the sense we have described: they
attempt to explain our observations by giving a broader,
presumably complete description of physical nature. The latter
confine themselves to give a set of logical rules regarding our
\emph{knowledge} of reality, putting aside the discussion of what
is \emph{really} happening. We find this kind of interpretations
unsatisfactory, and perhaps even opposed to the spirit of the
science of physics.

In this work we suggest a new ontological interpretation, which
attempts to overcome the difficulties mentioned. Namely it is
local, deterministic and simple, and it reconstructs the empirical
probability rules of standard quantum mechanics. The classical
properties mentioned in this section will be defined more
precisely and be further discussed in Chapter \ref{ch:prop}.

\section{Classicality and Decoherence}
\label{se:deco} The measurement problem, as it arises in a
situation where a classical apparatus is measuring a quantum
system, is tightly connected to the definition of the classicality
of systems. A classical system, by definition, is a system which
does not exhibit quantum-like behaviour in some sense. Generally,
we recognize one definite classical basis of states, of which our
classical system assumes one specific state. We do not observe
superpositions among classical states, in contrast to the case of
quantum states. Accordingly we may postulate that it is not
possible to interfere or mix the phases between two classical
states. In the case of a measurement process, it can be further
shown that without a scheme of choosing the classical basis and a
postulate such as above, it is not even well defined what
observable is being measured \cite{zurek81} \cite{zurek82}. This
is sometimes referred to as the problem of the preferred basis.

Selection of a classical basis and the destruction of phase
coherence can be achieved in the framework of standard quantum
mechanics, by the dynamical process of environment induced
superselection or decoherence, first introduced by Zurek
\cite{zurek81} \cite{zurek82} \cite{zurek91}. Here it is assumed
that the to-be-classical apparatus is being constantly monitored
by an environment, singling out an almost orthogonal basis. The
additional degrees of freedom, now coupled to the different
apparatus states in this basis, prohibit a change of this basis.
Doing so will result in the destruction of any previous
correlations of the apparatus with other systems, such as the
system being measured. Note that such a correlation is essential
for the definition of the classical basis. Due to the near
orthogonality of the environment states, when tracing out the
environmental degrees of freedom, one notices that the
off-diagonal interference terms in the reduced density matrix are
negligible. Therefore no projection onto a superposition of
classical states can take place. Here it is assumed that the
experimenter cannot take into account the environment
\emph{itself}, when performing such an experiment, since in
realistic situations the exact state of at least part of the
environment is unknown. Up to now we have not yet solved the
measurement problem -- no reduction to a \emph{single} classical
state has occurred, but importantly a basis for the reduction has
been defined.

The environment in discussion is the dominant part of the systems
``not of interest'', interacting with our apparatus. Usually these
may take the form of internal parts of our macroscopic device,
where the aforementioned ``apparatus'' is but the device's
pointer. The classical states of the preferred basis are called
``preferred states'' or ``pointer states''. As a result of the
diagonality of the density matrix, these states are insensitive to
measurements performed on them, in agreement with our classical
experience. By contrast, quantum states change or undergo
``preparation'' with each new measurement. Therefore we may relate
classicality to predictability. Zurek \cite{zurekpred} has even
suggested a predictability sieve, in order to identify the
classical states by the length of time that they maintain
predictability. With this in mind we shall later relate different
time measures to different measures of classicality of systems,
where for ideal classical systems, we would like the time they
remain predictable to be almost infinite or at least longer than
the lifetime of the universe.

To conclude the discussion we shall rederive a simple toy model of
decoherence presented in Ref. \cite{zurek82}. This model will be
later used also in the context of our interpretation. Let us start
with the initial state $a\uk+b\dk$ for the quantum system and
$\ak{R}=\hrt\sog{\ak{U}+\ak{D}}$ for the apparatus, both at time
$t_0=0$. Assume an interaction Hamiltonian between them of the
form
\begin{equation}
H_{sa}=-g\sig_z^{(s)}\ox\sig_y^{(a)},
\end{equation}
where $g$ is a constant, $\sig_y$ and $\sig_z$ are Pauli matrices,
and we write all states in the basis of eigenstates of $\sig_z$,
$(s)$ designates system and $(a)$ designates apparatus. Let the
interaction take place for a time
\begin{equation}
t_1=\frac{\pi\h}{4g}.
\end{equation}
The evolution will be \footnote{Derivations may be found in
Appendix \ref{ch:derivation}.}
\begin{equation}
\label{eq:s-a}
\hrt\sog{a\uk+b\dk}\ox\sog{\ak{U}+\ak{D}}\imp
a\uk\ox\ak{U}+b\dk\ox\ak{D}.
\end{equation}
Next take the state obtained and couple it to an environment
consisting of $N$ two-level particles with the basis $\{\ak{u}_k,
\ak{d}_k\}_{k=1}^N$, in the initial state
$\prod_{k=1}^N\ox\sog{\al_k\ak{u}_k+\bt_k\ak{d}_k}$. Assume an
interaction Hamiltonian between the apparatus and environment of
the form
\begin{equation}
H_{ae}=-\sig_z^{(a)}\ox\sum_{k=1}^Ng_k\sig_{z,k}^{(e)}\prod_{j\neq
k}\ox\mathrm{1}_j,
\end{equation}
where $g_k$ are constants, $(e)$ designates environment and
$\mathrm{1}$ is the unit operator. The evolution is
\begin{eqnarray}
\label{eq:a-e}
\lefteqn{\sog{a\uk\ox\ak{U}+b\dk\ox\ak{D}}%
\prod_{k=1}^N\ox\sog{\al_k\ak{u}_k+\bt_k\ak{d}_k}\imp}\nonumber\\
&&\imp
a\uk\ox\ak{U}\ox\ek{U}{t-t_1}+b\dk\ox\ak{D}\ox\ek{D}{t-t_1},
\end{eqnarray}
where {\setlength\arraycolsep{0pt}
\begin{eqnarray}
\ek{U}{t'}&=&\prod_{k=1}^N\ox\sog{\al_k\e{ig_kt'/\h}%
\ak{u}_k+\bt_k\e{-ig_kt'/\h}\ak{d}_k},\\
\ek{D}{t'}&=&\prod_{k=1}^N\ox\sog{\al_k\e{-ig_kt'/\h}%
\ak{u}_k+\bt_k\e{ig_kt'/\h}\ak{d}_k}
\end{eqnarray}}
are the environment states which correspond to the superselected
pointer states. Tracing out the environment, the reduced
system-apparatus density matrix obtained is
{\setlength\arraycolsep{0pt}
\begin{eqnarray}
\label{eq:rhod}
\rhod_{sa}(t)=\tr_e\rhod(t)&=&\abs{a}^2\uk\ub\ox\ak{U}\ab{U}+\nonumber\\
&+&z(t-t_1)ab^*\uk\db\ox\ak{U}\ab{D}+\nonumber\\
&+&z^*(t-t_1)ba^*\dk\ub\ox\ak{D}\ab{U}+\nonumber\\
&+&\abs{b}^2\dk\db\ox\ak{D}\ab{D}
\end{eqnarray}}
where
\begin{equation}
z(t')=\scal{\eps_U(t')}{\eps_D(t')}=\prod_{k=1}^N
\sog{\co{2g_kt'/\h}+i\sog{\abs{\al_k}^2-\abs{\bt_k}^2}\si{2g_kt'/\h}}
\end{equation}
is the correlation amplitude. The temporal average of the absolute
value of this amplitude is
\begin{equation}
\overline{\abs{z(t')}^2}=\lim_{T\to\infty}T^{-1}\int_0^T\abs{z(\tau)}^2d\tau=%
2^{-N}\prod_{k=1}^N\sog{1+\sog{\abs{\al_k}^2-\abs{\bt_k}^2}^2}.
\end{equation}
This implies that for large enough an environment, consisting of
many $N$ particles, $z(t)$ is negligible in average, and the
diagonal terms of (\ref{eq:rhod}), which contain it as a factor,
are damped out. Thus after some finite decoherence time, an
effective pointer basis has been established, and neither change
of basis nor projection onto a superposition of basis states are
possible. The only difficulty with this toy model is that since
$z(t)$ is of the almost-periodic function family, as long as $N$
is finite, any value of its range will recur an infinite number of
times \cite{recurrence}. Thus $z(t)$ will eventually return to
assume non-negligible values causing recoherence. Only when the
environmental degrees of freedom have a continuous spectrum of
eigenstates, can an infinitely long recoherence time be attained.

Classical systems should have a very short decoherence time and a
recoherence time longer than the lifetime of the universe. These
can be achieved when the environment is large, as indeed
characterizes measuring devices, which are macroscopic and
therefore have large environments. Of course a more realistic
model of an environment should be used. Since physical
interactions are usually a function of distance, given by a
potential, we would expect the pointer basis to be a position
basis, so that the different states are localized. However up to
now we have ignored the free Hamiltonian of the apparatus and
environment, assuming that they were commutative with the
interaction Hamiltonian. This is generally not true because the
momentum terms of the free Hamiltonians are not commutative with
the interaction potential. Thus a condition for localization is
massiveness of the apparatus \cite{dewitt}, which suppresses the
non-local terms. This again is true for macroscopic devices.

\chapter{Time Symmetric Quantum Mechanics}
\label{ch:symm}
\section{Background}
\label{se:background} In their famous 1964 article, ABL \cite{abl}
suggested a new rule for calculating probability. In the case that
a final state $\Psi_f$ is specified for the measured system, in
addition to the usual choice of an initial state $\Psi_i$, the
probability that an intermediate measurement of an operator $A$
yields the eigenstate $\ket{a_k}$ is {\setlength\arraycolsep{0pt}
\begin{eqnarray}
\label{eq:abl}
\prob(a_k\mid\Psi_i,\Psi_f)&=&\frac{\prob(\Psi_f(t)\mid a_k)\prob(a_k\mid\Psi_i(t))}%
{\sum_j{\prob(\Psi_f(t)\mid
a_j)\prob(a_j\mid\Psi_i(t))}}=\nonumber\\
&=&\frac{\abs{\scal{\Psi_f(t)}{a_k}}^2\abs{\scal{a_k}{\Psi_i(t)}}^2}%
{\sum_j\abs{\scal{\Psi_f(t)}{a_j}}^2\abs{\scal{a_j}{\Psi_i(t)}}^2},
\end{eqnarray}}
where $a_k\in\{a_j\}_j$, an eigenbasis of the measured operator
$A$, and we assume that an instantaneous measurement occurs at a
time $t$ intermediate of the boundary conditions, to which all
wave-functions are evolved. If the measurement is not
instantaneous, the initial and final wave-functions should be
taken at the beginning and ending of the measurement interaction,
respectively (assuming no free evolution of the wave-functions in
between).

If only the initial condition is specified, (\ref{eq:abl}) should
reduce to the regular empirical probability rule:
\begin{equation}
\label{eq:regprob}
\prob(a_k\mid\Psi_i)=\abs{\scal{a_k}{\Psi_i(t)}}^2.
\end{equation}
Formula (\ref{eq:abl}) is actually an application of the simple
conditional probability formula to the quantum case. But a
conceptual leap has been made in the recognition that boundary
conditions may be chosen time-symmetrically in contrast to the
conventional asymmetric choice of an initial condition only. The
gauntlet has been thrown down: why should boundary conditions be
chosen with such a discrimination?

Later Aharonov et. al. introduced the concept of weak measurement
and weak values \cite{spin100} \cite{weakval}. The idea is as
follows. If one performs an isolated, weak interaction
measurement, where the measured system is almost undisturbed and
no reduction takes place, his apparatus will show the expected
value of the measured operator, with a weighing of the appropriate
probabilities. A condition for this, as will be discussed in
Section \ref{se:weak}, is the existence of a large enough
uncertainty in the pointer's initial state. Then the pointer's
final state will be spread around the expected value, and the
final outcome observed by us, will be the result of an ideal
measurement on this spread state, with the appropriate probability
distribution.

When only an initial boundary condition is specified for the
quantum system, or when the final condition is identical to the
initial condition, one gets after interaction the expectation
value
\begin{equation}
\label{eq:expval} \langle
A\rangle\equiv\bra{\Psi_i}A\ket{\Psi_i}=\sum_ka_k\prob(a_k\mid\Psi_i).
\end{equation}
In the general case when both initial and final boundary
conditions are specified the outcome is the weak value $A_w$,
which may be far from any eigenstate of the measured operator
\begin{equation}
\label{eq:weakval}
A_w\equiv\frac{\bra{\Psi_f}A\ket{\Psi_i}}{\scal{\Psi_f}{\Psi_i}}\propto%
\sum_ka_k\sqrt{\prob(a_k\mid\Psi_i,\Psi_f)},
\end{equation}
where the times at which the wave-functions are taken are as
above. Notice that in (\ref{eq:weakval}) the appropriate weighing
of each eigenvalue is proportional to the \emph{square root} of
the ABL probability (\ref{eq:abl}).

The final boundary condition may result, for example, from
post-selection of the system after the interaction has taken
place, which can be achieved by performing an ideal measurement,
and discarding the cases with unwanted outcomes. Alternatively,
some systems in nature (as, we shall suggest, the universe) may
have an inherent final boundary condition, just as all systems
have initial ones. Weak values may play an important role in the
understanding of certain phenomena such as tunneling
\cite{tunneling} or Hawking radiation from a black hole
\cite{reznikrad}. Later, we shall consider weak measurements in
the frame of our interpretation.

In order to reduce the time-symmetric case to the
pre-selected-only case, one must choose the final state identical
to the initial state. This is immediately apparent when comparing
the weak value formula, left equation in (\ref{eq:weakval}), with
the expectation value formula, left equation in (\ref{eq:expval}).
A generalization to closed systems is achieved by taking the final
condition as the initial condition evolved to the final time, as
will be shown in Section \ref{sec:universe}. However when reducing
the ABL formula (\ref{eq:abl}) to the regular probability formula
(\ref{eq:regprob}), one must choose the final condition as the
measured $\ket{a_k}$, which is actually the evolved initial state,
assuming it has undergone reduction to a specific eigenstate. This
discrepancy arises due to the use of a probabilistic formula which
is foreign to our unitary formalism. This choice is also a clue
for the upcoming suggestion which will later justify it.

\section{The Two State Vector Formalism}
Following Ref. \cite{twostate}, we wish to reformulate quantum
mechanics, to be time-symmetric, in the sense that it will take
into account both initial and final boundary conditions. The
Schr\"odinger equation is first order in the time derivative,
therefore only one temporal boundary condition may be consistently
specified for a solution of the equation. Assuming both initial
and final boundary conditions exist, we must have two solutions
suitable for each of the two boundary conditions. The first is the
regular wave-function evolved forward in time from the initial
condition, which we call the ``history vector'' and denote by
$\ket{\Psi_\mathrm{HIS}(t)}$. The second is a \emph{different}
wave-function evolving from the future final condition, backwards
in time. We call the hermitian adjoint of this vector the
``destiny vector'', denoted by $\ket{\Psi_\mathrm{DES}(t)}$. We
postulate that the complete description of any system is given by
two vectors as such. These may be combined into operator form by
defining the ``two-state''
\begin{equation}
\rho(t)\equiv\frac{\ket{\Psi_\mathrm{HIS}(t)}\bra{\Psi_\mathrm{DES}(t)}}%
{\scal{\Psi_\mathrm{DES}(t)}{\Psi_\mathrm{HIS}(t)}}.
\end{equation}
This, in general, is not reducible to a single wave-function. It
is clear that orthogonal boundary conditions are forbidden,
therefore
\begin{equation}
\scal{\Psi_\mathrm{DES}(t)}{\Psi_\mathrm{HIS}(t)}\ne0,
\end{equation}
which is a reasonable choice due to the fact that a final state,
orthogonal to the initial state, has probability zero for being
post-selected. For a given Hamiltonian $H(t)$, the time evolution
of the two-state from time $t_1$ to $t_2$ is
\begin{equation}
\rho(t_2)=U(t_2,t_1)\rho(t_1)U(t_1,t_2),
\end{equation}
where $U(t_2,t_1)$ is the regular evolution operator
\begin{equation}
U(t_2,t_1)=\e{-i/\h\int_{t_1}^{t_2}H(\tau)d\tau}.
\end{equation}
The two-state takes the place of the density matrix in standard
quantum mechanics. Any subsystem's two-state may be obtained by
taking the partial trace of all other degrees of freedom. In the
current work, our formalism supports only these two operations:
unitary time evolution and tracing. No other formula is allowed
nor required, as we shall show in the next chapter.

\section{The Two-State of the Universe}
\label{sec:universe} The Conventional approach to nonrelativistic
quantum mechanics assumes that a complete description of the state
of a closed system, such as the universe, is given at any time $t$
by a wave-function
\begin{equation}
\ket{\Psi(t)}=U(t,t_0)\ket{\Psi(t_0)},
\end{equation}
where $\Psi(t_0)$ is usually defined at some initial time
$t_0=t_i$ where $t>t_i$, hence $\Psi(t_0)$ is the initial boundary
condition $\Psi_i(t_i)$. We shall denote this wave-function
$\Psi_i$. The density matrix associated with the state
$\ket{\Psi_i(t)}$ is
\begin{equation}
\rhod(t)=\ket{\Psi_i(t)}\bra{\Psi_i(t)},
\end{equation}
and the state of any subsystem is obtained by taking the partial
trace of all other degrees of freedom.

We will show that the above description is consistent with the
assumption of a special final boundary condition of the form
\begin{equation}
\label{eq:f=i} \ket{\Psi_f(t_f)}=U(t_f,t_i)\ket{\Psi_i(t_i)},
\end{equation}
at the final time $t_f$, for a backward evolving wave-function
denoted $\Psi_f$. This will be established by writing the
two-state at any intermediate time. Up to a normalization factor
{\setlength\arraycolsep{0pt}
\begin{eqnarray}
\rho(t)&=&\ket{\Psi_i(t)}\bra{\Psi_f(t)}=\nonumber\\
&=&\ket{\Psi_i(t)}\bra{\Psi_f(t_f)}U(t_f,t)=\nonumber\\
&=&\ket{\Psi_i(t)}\bra{\Psi_i(t_i)}U(t_f,t_i)U(t_f,t)=\nonumber\\
&=&\ket{\Psi_i(t)}\bra{\Psi_i(t_i)}U(t_i,t)=\nonumber\\
&=&\ket{\Psi_i(t)}\bra{\Psi_i(t)}=\rhod(t).
\end{eqnarray}}
Therefore if only an initial boundary condition is assumed, in all
calculations two-states may by substituted by density matrices,
giving the same results. In the previous section's notation we
have simply taken the destiny vector to be equal to the history
vector.

Under the assumption of a deterministic evolution rule for the
universe, which will be justified later, we have shown that our
formalism is reducible to the conventional one, the latter being a
special case of the former. Taking final boundary conditions
different from (\ref{eq:f=i}), our formalism introduces a richer
state structure into quantum theory. This is a generalization, to
the closed universe, of the ABL suggestion, of choosing two
temporal boundary conditions for the system being measured.

In the next chapter we suggest postulating a very special final
boundary condition at the final time, a time which should be as
late as the lifetime of the universe, whether the universe ends in
a singularity, in relaxation to a steady state, or is infinite in
time.

\chapter{Teleological Interpretation of Quantum Mechanics}
\label{ch:tele}
\section{The Suggestion}
We are now ready to present our suggestion for a new
interpretation of quantum mechanics, the \emph{Teleological
Interpretation}. The Webster New Collegiate Dictionary defines
``Teleology'' as
\begin{quotation}
1 a: the study of evidences of design in nature b: a doctrine (as
in vitalism) that ends are immanent in nature c: a doctrine
explaining phenomena by final causes 2: the fact or character
attributed to nature or natural processes of being directed toward
an end or shaped by a purpose 3: the use of design or purpose as
an explanation of natural phenomena.
\end{quotation}
We argue that special final conditions may exist, so that if they
are taken into account, the probabilistic predictions of quantum
mechanics may be explained. Setting aside for the moment the
mechanism by which a proper chosen final state causes an effective
reduction to the appropriate desired outcome, we wish to present
the scheme by which the final states should by selected. We argue
that a universe set up as follows behaves as predicted by standard
quantum mechanics, such as we believe our universe does, using
only unitary Schr\"odinger evolution.
\begin{itemize}
\item Choose the universe's initial boundary condition.
\item Choose the universal Hamiltonian.
\item Identify the classical systems and their preferred basis.
\item Identify measurement-like interactions between classical and
quantum systems.
\item Assign the universe's final boundary condition as the initial one, evolved to the final
time, with the following exceptions:
\begin{itemize}
\item For classical systems select one of the preferred basis
states (normalized) from the superposition, while ensuring that
the probability distribution for measurements on large ensemble
match the one predicted by the regular probability rules.
\item Set special final states to produce strange phenomena (optional).
\item Set special final states to match the free will of sentient beings (optional).
\end{itemize}
\end{itemize}
Assuming such boundary conditions for the universe, ideal
``probabilistic'' and non-ideal weak measurements will be analyzed
in the following sections. Other than determining the preset
distribution, the probability rules mentioned in Section
\ref{se:background} need not to be applied in any manner, on the
contrary, we treat them as empirical rules and show how their
predictions are reconstructed. Therefore no ``projections'' should
be applied to the system during the measurement process. We must
stress that in most situations the final classical states are not
known to us prior to the completion of the measurement, and are
assumed to be such or the other for the sake of the computational
examples. The reader should not concern himself at this stage with
questions of the amount of freedom of choice in performing
different measurements. We will later examine the applicability of
the concept of free will in the framework of the suggested
interpretation.

\section{Ideal Measurements}
\label{se:ideal} In this section we shall analyze the ideal
measurement process, and show how effective reduction takes place.
For the sake of simplicity, we stick with the decoherence toy
model presented in Section \ref{se:deco}. Although simple, it is
important to take such a \emph{dynamic} model in order to fully
understand the relevant processes. There is some resemblance to
the work done in Ref. \cite{reznikenv}. Recall the state obtained
in Section \ref{se:deco} after decoherence. Let us take this state
as our history vector
\begin{equation}
\ket{\Psi_\mathrm{HIS}(t)}=a\uk\ox\ak{U}\ox\ek{U}{t-t_1}+b\dk\ox\ak{D}\ox\ek{D}{t-t_1},
\end{equation}
where $t_1$ was some finite system-apparatus interaction time and
$\ak{U}, \ak{D}$ were superselected as the classical pointer
states. Let us assume that the environment associated with these
states is large enough so that the recoherence time, or the time
which takes the correlation amplitude to revert to a
non-negligible value, is longer than the lifetime of the universe.
It is then possible to assign the final boundary condition as
\begin{equation}
\ket{...}\ox\ak{U}\ox\ek{U}{t_f},
\end{equation}
for example, where $\ket{...}$ are unknown systems correlated to
our pointer state, and a specific state was chosen for the pointer
$\ak{U}$, from its classical basis, correlated to the adequate
environment state of that time $\ek{U}{t_f}$. Such a choice is
reasonable because after decoherence has taken place and before
recoherence, no interference between the pointer states can take
place anyway. The quantum system, by contrast, has a certain
definite state only until the next measurement made on it, which
prepares it in a new state
\begin{equation}
\ket{\phi}=c\uk+d\dk,
\end{equation}
which takes the role of an effective final boundary condition for
the quantum system as will be soon showed. Therefore the
\emph{destiny vector} at a time after the measurement interaction
is over is
\begin{equation}
\ket{\Psi_\mathrm{DES}(t)}=\ket{\phi}\ox\ak{U}\ox\ek{U}{t-t_1}.
\end{equation}
The complete description of our systems is given up to a
normalization factor by the two-state {\setlength\arraycolsep{0pt}
\begin{eqnarray}
\label{eq:rho}
\rho(t)&=&\ket{\Psi_\mathrm{HIS}(t)}\bra{\Psi_\mathrm{DES}(t)}=\nonumber\\
&=&a\uk\ox\ak{U}\ox\ek{U}{t-t_1}\bra{\phi}\ox\ab{U}\ox\eb{U}{t-t_1}+\nonumber\\
&+&b\dk\ox\ak{D}\ox\ek{D}{t-t_1}\bra{\phi}\ox\ab{U}\ox\eb{U}{t-t_1}.
\end{eqnarray}}
Ignoring the environment the reduced two-state obtained is
{\setlength\arraycolsep{0pt}
\begin{eqnarray}
\rho_{sa}(t)=\tr_e\rho(t)&=&a\uk\bra{\phi}\ox\ak{U}\ab{U}+\nonumber\\
&+&z^*(t-t_1)b\dk\bra{\phi}\ox\ak{D}\ab{U}.
\end{eqnarray}}
Refer to Section \ref{se:deco} to recall the behaviour of $z(t)$.
It is evident that after a decoherence time an effective reduction
of the pointer to the state: $\ak{U}\ab{U}$ has occurred. We call
this process ``two-time decoherence'' differing it from the
regular meaning of decoherence.

While the above is true for the pointer involved in the original
measurement, notice that the state of the quantum system and the
state of any apparatus performing \emph{further} measurements in
the same basis of the quantum system or original pointer,
experience immediate effective reduction, with no decoherence time
delay. This becomes evident, after tracing out the \emph{original
pointer's} degree of freedom, as in the above example, the
off-diagonal term which contained $\ak{U}\ab{D}$ vanishes in their
reduced two-states. In the ``old'' probabilistic nomenclature,
applying the ABL rule, the original pointer has probability $1$ to
be found (by those correlated systems) in the reduced state, and
probability $0$ to be found in any other state. Therefore in a
realistic experiment where a chain of measurements exists, one
should expect to observe immediate reduction, in contrast to the
prediction of the many worlds interpretation, for example, which
states that a decoherence time until effective reduction, should
always be expected. This may be an important deviation point when
comparing which of the two interpretations is more applicable. It
may be that the two-time decoherence does not play an important
role in the measurement chain, but as we have shown, it naturally
emerges from the combination of regular decoherence needed for
classicality, and the two-state formalism with our special final
boundary condition. Of course it remains essential that the final
state is chosen from the classical pointer states superselected by
regular decoherence. In the many worlds picture where each
superposition term is viewed as a branching world, we have simply
selected one specific branch. Such a view may help seeing why the
interpretation suggested is self consistent, and why is it
naturally demanded, if one does not wish to have a multitude of
``worlds''.

It remains to show how the effective reduction determines the
backward evolving state of the quantum system, for a previous
measurement, as the process presented above shows effective
reduction only after a finite positive time duration of the
system-apparatus interaction. Evolving the destiny vector at time
$t_1$, $\sog{c\uk+d\dk}\ox\ak{U}$, backwards to the time $t_0=0$
(which was chosen for convenience as the beginning of the
measurement), the two-state of the quantum system and apparatus
obtained is (up to normalization){\setlength\arraycolsep{0pt}
\begin{eqnarray}
\rho_{sa}(0)&=&ac^*\uk\ub\ox\ak{R}\ab{R}+ad^*\uk\db\ox\ak{R}\ab{L}+\nonumber\\
&+&bc^*\dk\ub\ox\ak{R}\ab{R}+bd^*\dk\db\ox\ak{R}\ab{L},
\end{eqnarray}}
where $\ak{R}=\hrt\sog{\ak{U}+\ak{D}}$ and
$\ak{L}=\hrt\sog{\ak{U}-\ak{D}}$. Taking the partial trace on the
apparatus' degree of freedom, shows that the backward evolving
vector of the quantum state is $\ub$ as expected from this
process, in which the outcome was ``UP''. This sets a final
boundary condition for the quantum state in a previous
measurement, in the same manner that we have taken into account
the state $\ket{\phi}$ from the next measurement.

We have shown how effective reduction may take place in an ideal
measurement when the final state of the classical apparatus is
chosen as one of its possible classical states after the
measurement. Setting a final boundary condition for the classical
states at a \emph{very} late time, enables them to stay
predictable as expected. Conversely, the quantum system is
``prepared'' at each measurement in a new state which constitutes
an effective boundary conditions for both the next and previous
measurement. In our analysis we have treated these as our initial
and final boundary conditions for simplicity, bringing the final
apparatus state from the absolute final time, under the assumption
that it has not undergone any further interaction. We have thus
formulated a connection between the classicality of a system and
the length of time between its initial and final boundary
conditions.

Our example considers only a single measurement process per
measuring device. If the apparatus undergoes multiple
interactions, always some initialization process of the measuring
device must take place, one which is dependent on the previous
measurement's outcome. This is obvious, due to the fact that
information cannot be lost but is always transferred to other
systems. Hence, the information of most of the measurements'
outcomes reside in those systems' final state.

\section{Weak Measurements}
\label{se:weak} We now wish to analyze non-ideal weak
measurements, which may be naturally described using the two-state
formalism. A weak measurement is one in which the precision of the
measurement is low enough, so that negligible change would be
induced to the measured system. In a weak measurement the
measuring device gets correlated to the different eigenstates of
the system, but no reduction to a specific eigenstate takes place.
What is measured is an average ``weak'' value of the operator,
which is dependent on the initial and final states of the system.
Let us take a many level quantum system in a basis of the states
$\ket{a_k}$ on which the operator $A$ is defined as
\begin{equation}
A=\sum_ka_k\ket{a_k}\bra{a_k}.
\end{equation}
The initial state of the system is chosen to be
\begin{equation}
\ket{\phi_1}=\sum_kc_k\ket{a_k},
\end{equation}
where $c_k$ are constants. Next we take the measuring device as a
pointer, with its position $q$ described by a Gaussian-like
function $Q(q)$, initially set as $Q(0)$. We let an interaction
between the system and apparatus take place under the Hamiltonian
\begin{equation}
H=-g(t)PA,
\end{equation}
until the time $t_1$, where
\begin{equation}
\int_0^{t_1}g(\tau)d\tau=1,
\end{equation}
and $P$ is the operator of the momentum conjugate to $q$. The
evolved state of the apparatus after the time $t_1$ is given by
the law:
\begin{equation}
\e{ia_kP/\h}\ket{Q(0)}=\ket{Q(a_k)},
\end{equation}
for eigenvalues $a_k$ of $A$. Let us take the initial state of our
composite system to be
\begin{equation}
\ket{\Psi_i(0)}=\ket{\phi_1}\ox\ket{Q(0)}.
\end{equation}
At the time $t=t_1$ the evolved initial state is
\begin{equation}
\ket{\Psi_i(t_1)}=\sum_kc_k\ket{a_k}\ox\ket{Q(a_k)}.
\end{equation}
At some time $t_2>t_1$ we perform an ideal measurement on the
quantum system and obtain the result
\begin{equation}
\ket{\phi_2}=\sum_kc_k'\ket{a_k},
\end{equation}
which serves as a final boundary condition for the quantum system
as explained in the previous section. A calculation shows that the
final composite state must then be
\begin{equation}
\ket{\Psi_f(t_2)}=\ket{\phi_2}\bra{\phi_2}\ox\ket{\Psi_i(t_1)}=%
\ket{\phi_2}\ox\sum_jc_jc_j'^*\ket{Q(a_j)}.
\end{equation}
The two-state at a time $t$ between $t_1$ and $t_2$ (after
interaction and before post-selection, and assuming no free
evolution) is given up to a normalization factor by
\begin{equation}
\rho(t)=\ket{\Psi_i(t_1)}\bra{\Psi_f(t_2)},
\end{equation}
and the apparatus' pointer will show
\begin{equation}
\rho_a(t)=\tr_s\rho(t)=\sum_{k,j}c_kc_k'^*c_j^*c_j'\ket{Q(a_k)}\bra{Q(a_j)}.
\end{equation}
The condition for weakness of the measurement is that the
Gaussians are wide enough so that the relation
\begin{equation}
\sum_kc_kc_k'^*\ket{Q(a_k)}\cong\sum_kc_kc_k'^*\ket{Q(a')}\equiv\ket{\hat{Q}(a')},
\end{equation}
for some $a'$, holds due to their interference. Then tracing out
the quantum system's degree of freedom, the apparatus reads
\begin{equation}
\rho_a(t)\cong\ket{\hat{Q}(a')}\bra{\hat{Q}(a')},
\end{equation}
where it is shown that $a'$ equals $A_w$, the weak value of $A$,
\begin{equation}
A_w=\frac{\bra{\phi_1}A\ket{\phi_2}}{\scal{\phi_1}{\phi_2}}=%
\frac{\sum_kc_kc_k'^*a_k}{\sum_kc_kc_k'^*},
\end{equation}
by computing the weak value of the evolution operator, as follows:
{\setlength\arraycolsep{0pt}
\begin{eqnarray}
\sum_kc_kc_k'^*\ket{Q(a_k)}&=&\sum_kc_kc_k'^*\e{ia_kP/\h}\ket{Q(0)}=\nonumber\\
&=&\bra{\phi_2}\e{iAP/\h}\ket{\phi_1}\ket{Q(0)}=\nonumber\\
&=&\sum_{n=1}^\infty\frac{\sog{iP/\h}^n}{n!}\bra{\phi_2}A^n\ket{\phi_1}\ket{Q(0)}=\nonumber\\
&=&\scal{\phi_2}{\phi_1}\e{iA_wP/\h}\ket{Q(0)}+\nonumber\\
&+&\sum_{n=2}^\infty\frac{\sog{iP/\h}^n}{n!}%
\sog{\bra{\phi_2}A^n\ket{\phi_1}-\bra{\phi_2}A\ket{\phi_1}^n}\ket{Q(0)}\cong\nonumber\\
&\cong&\scal{\phi_2}{\phi_1}\e{iA_wP/\h}\ket{Q(0)}=\ket{\hat{Q}(A_w)},
\end{eqnarray}}
where we have applied the usual condition for weakness
\cite{spin100} \cite{weakval}, which requires that the terms with
$n>1$ in the Taylor expansion are negligible, given the choice of
a wide enough initial Gaussian.

It is clear that after the post-selection takes place, the weak
value emerges as a consequence of a projection onto the new
quantum state. The weak value actually exist also \emph{before}
the post-selection, after the system-apparatus interaction is
completed. This fact is usually explained by the reasoning that
the order of actions, looking at the pointer and performing the
post-selection, is unimportant, as the post-selection cannot
affect the measuring device after the measuring interaction is
over. Thus the apparatus must show the same value even before
post-selection takes place. In the above two-state formulation,
the weak measurement is clearly seen to arise before the
post-selection, when looking at the reduced two-state of the
apparatus after tracing out the quantum system.

If the weakness condition is not satisfied but the measurement is
not an ideal one, we are dealing with a regime of measurement of
intermediate strength. In some of these cases, the outcome of the
measurement may be given by a combination of the ideal and weak
mechanisms, as the outcome of an ideal measurement on a set of
different weak values \cite{twostate}.

Measurements performed by us, on large, uncontrollable systems,
may satisfy the weakness condition. Measurements of galactic
properties which yield too large a spin or magnetic moment or
mass, may be due to a special final boundary condition for that
stellar object, which yields a weak value, far from the expected
eigenvalue, calculated by theoretical means. The problem of the
missing mass, a recent discovery that the universe seems too young
or more generally inconsistencies in measurements of the
cosmological constants by different methods \cite{cosmo}, may as
well be explained by a special final boundary condition for the
universe. Another example may be the observation that the
calculated number of Darwin mutations seems to be too low to
explain the genetic evolution of complex life forms. Perhaps these
sorts of strange phenomena may be explained by assuming the
existence of special final boundary conditions for these systems,
which would appear to us as new fundamental laws of our universe.
Of course when dealing with everyday low-energy short-duration
experiments, these should reduce to give the expected regular
results \cite{reznikrad}. It might trouble some physicists, or
please others, if such a special boundary condition could be used
to break the restrictions of causality.

\chapter{The Classical Properties}
\label{ch:prop}

\section{Causality, Locality and Realism}
In this discussion we refer to ``causality'' as the impossibility
of superluminal signaling or of advanced action. In the context of
quantum theory this means that probability distribution of
experiments' outcomes cannot be affected by events outside of
their past light cone. We use ``locality'' for the stronger
property of complete prohibition of any action at a distance or
advanced action, meaning that there may be no influence on any
events outside the future light-cone, including the quantum state
itself, even if it does not imply defiance of causality. The
second property (and of course the first, being implied by it)
should exist in order to retain consistency with the theory of
relativity, which we axiomatically assume holds.

The standard quantum theory does not contradict the concept of
causality in any relativistic or nonrelativistic setup
\cite{nosuper1} \cite{nosuper2} \cite{nosuper3}. We wish to show
that causality is maintained also in the framework of the
suggested interpretation. Causality requires ignorance of the
future boundary state, if such exists, for knowledge of it would
allow the existence of the above restricted effects. It is to be
remembered that in all examples of the previous chapters, the
final boundary state was assumed to take a specific state, for the
sake of demonstration, and generally, it cannot be known a-priori.
In fact only when an second identical measurement is performed in
sequence to the preparation measurement, can the final state be
known for certain (``with probability $1$''). In the discussion of
ideal measurements in Section \ref{se:ideal}, we have shown that
our interpretation is consistent with the conventional formalism
and therefore causal. Neither can weak measurements defy
causality, because, due to the weakness condition, there is large
uncertainty in the outcome of the weak measurement, and it may
yield values in the proximity of a strange value, appropriate to
some final condition, also for other final conditions. In fact, an
error as such will result in the majority of cases. Therefore it
is not possible to deduce from the outcome of a weak measurement
on the final state.

In order to demonstrate how forbidden knowledge of a final
boundary state (a non-causal state of affairs by itself) could
have been used for superluminal signaling, examine the following
setup. We work with two spin-$\half$ particles located at two far
away locations. We start with the initial state
$\hrt\sog{\uk_L\ox\uk_R+\dk_L\ox\dk_R}$, and assume we know the
final state to be $\hrt\sog{\uk_L\ox\uk_R+\uk_L\ox\dk_R}$, where
$L$ denotes the left particle and $R$ the right particle. An
observer on the left may or may not perform a unitary rotation on
his particle of the form $\uk_L\imp\dk_L$ and $\dk_L\imp\uk_L$,
leaving the initial composite state as it is or transforming it
into the state $\hrt\sog{\dk_L\ox\uk_R+\uk_L\ox\dk_R}$. Now the
observer on the right measures the spin of his particle, obtaining
$\dk_R$ or $\uk_R$, according to the action or non-action of his
friend on the left. In this manner, the left observer may
allegedly transmit signals to the right observer in an instant. A
procedure like this would be possible for many arbitrary choices
of initial and final states.

The theory of relativity states that physics is simple only when
analyzed locally. Therefore we would like it to be \emph{possible}
to analyze physics locally. Moreover non-local effects are strange
to the theory of relativity as they defy Lorentz covariance.
Quantum theory, due to its linear structure, has an inherent
non-locality encompassed in its entangled states (a spatially
separated correlated pair, for example). This property is
sometimes called ``wholeness'' or ``inseparability'' of quantum
mechanics. This non-locality might raise causality concerns when
performing measurements on part of the entangled state. A local
interpretation of quantum mechanics would have to show how could
the outcomes of such a measurement be explained locally. Take for
example the non-local collapse interpretation; the collapse of the
wave-function is assumed to take place instantaneously in all
space. It is well known that such a phenomenon cannot be used for
superluminal signaling, as mentioned above. But even so, the mere
concept of instantaneous state collapse is non-covariant, for even
if the collapse events are simultaneous in some space-time
hyper-surface, these events would not occur at the same time on
any other hyper-surface when changing the frame of reference
\cite{noncovar}. This is a common pitfall of collapse
interpretations. On the other hand, the picture presented by the
interpretation suggested by us, gives a purely local explanation
for the effective state reduction.

The property of ``realism'' or ``objectivity'', represents the
classical concept that physical nature exist independently of
possible observations of it. The lack of covariance of the
collapse process is an example in which the description of nature
is dependent on the observer. EPR \cite{epr} define realism by the
following counterfactual:
\begin{quotation}
If, without in any way disturbing a system, we can predict with
certainty (i.e., with probability equal to unity) the value of a
physical quantity, then there exists an element of physical
reality corresponding to this physical quantity.
\end{quotation}
This means that a (possibly ``hidden'') variable exists, which
determines the outcome of the measurement. The concern of whether
this variable is actually accessible to us, goes to the discussion
of free will in the following section. As mentioned before in
Section \ref{se:inter} (\cite{bell},\cite{ghz},\cite{mermin}) it
has been proved that locality and realism cannot coexist in any
interpretation of quantum mechanics. This proof rests on the
supplemental assumptions that no non-causal action is allowed,
that asking questions of ``what if'' is permitted (there is
``counterfactual definiteness'' \cite{cfd}) and that there are no
predetermined ``conspiratorial'' dependencies between what has
been measured and what has outcome. Then no local hidden variables
theory can exist, and at least one of these concepts should be
given up. In the many worlds interpretation counterfactual
definiteness does not hold, because of the coexistence of multiple
outcomes for a measurement, therefore the discussion of realism is
irrelevant. Also in our interpretation, realism is irrelevant,
because we have assumed predetermined final boundary conditions
which determine the outcomes of measurements, knowing from the
initial state and Hamiltonian what observations actually take
place. Bell described such a situation as apparently separate
parts of the world being deeply and conspiratorially entangled.

\section{Determinism and Free Will}
The classic property of ``determinism'' signifies the ability to
completely deduce the state of the system at any time, only from
the knowledge of its initial boundary condition and evolution
rules. Sometimes this is also referred to as ``causality'' but we
reserve the latter term for the relativistic meaning discussed
above. The concept of determinism constitutes a main disagreement
point between the different interpretations of the quantum theory.
We have addressed deterministic and nondeterministic
interpretations in Section \ref{se:inter}. Deterministic
interpretations strive to reach a complete description of reality
with the introduction of additional rules into standard quantum
mechanics, in order to make the behaviour of any system
predictable \emph{in principle}, as opposed to the random collapse
mechanism. This is in the spirit of Einstein's famous saying that
``God does not play dice''. We claim that our suggested
interpretation is deterministic in a broader, two-time sense,
where the given boundary conditions also include the final
condition. With this, specific outcomes of measurements are
completely accounted for, while the probabilistic structure arises
due to a preset stochastic distribution in the final boundary
condition, and the usual lack of knowledge of the specific final
states therein. The latter is a requirement of causality as
discussed before.

The discussion of free will or freedom of choice may seem awkward
in the scope of a thesis in physics, but since these concepts have
considerably bothered many philosophers (including the author) in
their attempt to explain reality, we find the discussion relevant
in the context of a theory which strives to give a complete
description of reality. We make a discrimination between effective
or apparent free will which means that any specific observer may
effectively experience freedom of choice, and essential or real
free will which is more a moral concept, of whether an individual
is ``really'' free, in some sense, to choose his course of action.

We say that a system has an effective free will, or is a ``free
agent'', if at some instant a choice can be made between different
possible evolutions, independently of any accessible past data, or
in short, if there is freedom from the past. A theory which allows
the existence of free agents must include a free-from-the-past
mechanism which can supply different outcomes, as a function of
something different from accessible past variables. For example,
if there had existed a mechanism which generates purely random
numbers, it would have served this purpose adequately. The
empirical probabilistic behaviour of quantum mechanics, never mind
the interpretation explaining it, would do as well as a mechanism
which permits effective freedom of choice. Notice that the
effective free will experienced by us, Humans, may well be due to
our lack of complete knowledge of the exact past or current state
of the relevant systems (intensified by chaotic processes) and it
may even be to some extent a psychological illusion. By stating
that a theory allows effective free will, we permit but not prove,
Humans to be free agents. The possibility of effective free will
is implied by, but does not necessarily imply, lack of
determinism, for we have only required independence of accessible
past data, while also a deterministic theory such as hidden
variables can supply the necessary free-from-the-past mechanism,
as the hidden variables determine the outcome of the measurement,
but are themselves inaccessible. We have further broadened the
concept of determinism to be time-symmetric, where for free will
we do not do so, because the future data is anyway inaccessible to
us, due to the restrictions of causality.

The possibility of effective freedom of choice may be demonstrated
in our interpretation as follows. We analyze two large macroscopic
isolated systems, such as two distant galaxies. System A
containing Humans and system B containing aliens. Assume that in
system A the natural time flows toward the future as usual, while
in system B it flows backwards towards the past. This may be
achieved for system B by choosing a special initial boundary
condition such that the entropy decreases with the usual flow of
time as it is in system A; then in system B the thermodynamic
arrow of time is reversed in respect to system A. Assume that both
systems are classical in the sense that system B is monitoring
system A without disturbing it. Now the past of system B
encompasses the future state of system A. Call these data the
future boundary condition of system A. As long as system B does
not pass any information to system A (an action which would cause
severe causal problems), system A cannot purposely change its
choice to be inconsistent with (and thus spoil) system B's memory.
Now let system A be quantum mechanical. Dependence of events on
the past is not obliged because of the probabilistic structure of
quantum mechanics, as explained above. In our interpretation this
is expressed in the unknown final boundary condition chosen for
the universe. This picture perhaps suits an old saying of
\emph{Our Sages} in \emph{Pirkei Avot}, Chapter 3, Verse 19: ``All
is foreseen and the choice is given''.

Moving on to essential free will; Humans have such a strong
tendency to believe in their freedom of choice, that whole moral
doctrines rely upon it. We believe that the individual chooses how
to act (for example whether to be good or evil) and therefore he
is responsible for his deeds and influences his fate. We have seen
that systems may well effectively behave as having free will. But
when analyzing the origin of a choice made, after discarding all
past dependencies one is left only with the quantum mechanic
probabilistic structure which supplies a stochastic
free-from-the-past mechanism. How can such a mechanism reflect an
individual's free sober choice? Even if free from the past,
effective freedom of choice is not a real choice of \emph{the
individual}, and the essence of the concept is lost. Actually it
is quite difficult to think of any theoretical model which would
allow such a thing. Now the question should be asked, whether we
believe in the concept of free will strongly enough to
\emph{demand} such a theory. If so, we would like the deviation
required from our current physical theory to be minimal. It may be
that our interpretation could meet these demands. We have
mentioned that effective freedom of choice is due to the
probabilistic structure which is embodied in the final boundary
condition in our interpretation. If we assume that for some reason
the states of this boundary condition are somehow compatible with
the volitional choices of Human beings, we can achieve a mechanism
for essential free will. Recalling that the data in the final
boundary condition are inaccessible, we are self consistent as in
the example above. Alternatively one may choose to look at things
in the following manner. The complete description of any system
encompasses both a history vector and a destiny vector, making the
future also part of the system. Discarding its past and demanding
the choice to be made by \emph{the system}, one is left only with
its \emph{future}, to determine the choice. This might suggest
that a new conception of time, should be adopted: one in which our
treatment of time is more complex than the usual one-dimensional
notion. Such an approach might be necessary when one attempts to
describe the logical picture of events, since the cause for
actions of systems comes from both their past and their future,
while the effect occurs at some present time. Perhaps the words of
T. S. Eliot in ``Four Quartets'' are suitable for this picture:
\begin{verse}
Time past and time future\\ What might have been and what has
been\\ Point to one end, which is always present.
\end{verse}

\chapter{Conclusion}
\label{ch:last}
\section{Discussion}
It is interesting to compare our suggested interpretation to
similar interpretations of quantum mechanics and to related
philosophical ideas in general. In Section \ref{se:ideal} we have
already made some comparisons to the many worlds interpretation.

The idea of applying quantum mechanics to the whole universe and
discussing its wave-function, and adequate boundary conditions,
was introduced by Hartle and Hawking \cite{harthawk}.

The two-state formalism may be extended to a formalism of multiple
time boundaries \cite{twostate}. Also, consistent history
interpretations \cite{histories89} \cite{histories90} are based on
a history of projections. As we have shown, such an approach is
unnecessary in order to solve the measurement problem.

In the course of this work we have mainly used the formalism of
two-states for its formal convenience, although as stated before,
one may look at the complete state of the system as constituting
the regular history vector, and of a destiny vector which
determines the ``fate'' of the system. This recalls the ideas of
the philosopher Henry Bergson, of systems having some internal
tendency, or inner motive, ``\'elan vital'', towards a certain
destiny.

We have not yet considered the question of the role of the quantum
mechanical measurement process in determining the arrow of time.
We rely on decoherence as the cause of irreversible singling of
the classical basis of states for classical systems, which lasts
until the late end of the universe. Only then do we allow a final
boundary condition which causes effective reduction. Hence we do
not increase the measure of irreversibility beyond that created by
the thermodynamic arrow of time (which itself is usually assumed
to be a consequence of the cosmological arrow of time). Therefore
it seems that in our interpretation there is no microscopic
quantum mechanical reason for irreversibility, nor does our choice
of the special final boundary condition increase the
irreversibility determined by the initial condition of the
universe.

\section{Summary}
We have suggested a new interpretation of quantum mechanics, the
\emph{Teleological Interpretation}. We have shown that a special
final boundary condition may be chosen, one which causes effective
reduction consistent with the predictions of standard quantum
mechanics, thus solving the measurement problem. Because the
deviation taken from the conventional theory is minimal, we dare
to state that by Occam's principle, it may well be that such a
final boundary condition indeed exists for our universe.

We think that the suggested interpretation may answer some of the
problems and gaps left by previous interpretations of quantum
mechanics. We hope that this work will be the beginning of an
adequate response to the many articles discussing the
time-symmetric formulation of quantum mechanics, which end with
the statement, that the formulation may lead to a new
interpretation of quantum mechanics.

\section{Acknowledgments}
The Author would like to thank his instructors: Prof. Yakir
Aharonov and Prof. Issachar Unna, for their inspirational guidance
of the work. Also much gratitude goes to Dr. Lev Vaidman, to Dr.
Benny Reznik, and to Dr. Daniel Rohrlich for reviewing and
commenting on this work and for many helpful conversations.
\appendix
\chapter{Derivation of Discussed States}
\label{ch:derivation}
The following trivial derivations are are
given for the completeness of this work. We use the identities:
\begin{displaymath}
\e{ic\sig_z}=\sog{\begin{array}{cc}%
\e{ic} & 0\\%
0 & \e{-ic}
\end{array}},
\end{displaymath}
\begin{displaymath}
\e{ic\sig_y}=\sog{\begin{array}{cc}%
\co{c} & \si{c}\\%
-\si{c} & \co{c}
\end{array}},
\end{displaymath}
where $c$ is a constant, $\sig_y$ and $\sig_z$ are Pauli matrices.
The designations $(s)$, $(a)$ and $(e)$ stand for system,
apparatus and environment respectively. $g$ and $g_k$ are
constants. Derivation of (\ref{eq:s-a}):
{\setlength\arraycolsep{0pt}
\begin{eqnarray*}
&&H_{sa}=-g\sig_z^{(s)}\ox\sig_y^{(a)}.\\
&&\ket{\Psi_{sa}^i(0)}=\hrt\sog{a\uk+b\dk}\ox\sog{\ak{U}+\ak{D}}.\\
&&\ket{\Psi_{sa}^i(0\leq t\leq
t_1)}=\hrt\sog{{\setlength\arraycolsep{5pt}\begin{array}{cc}%
\e{ig\sig_y^{(a)}t/\h} & 0\\ %
0 & \e{-ig\sig_y^{(a)}t/\h}
\end{array}}}_s\sog{\begin{array}{c}%
a\\%
b
\end{array}}_s\ox\\
&&\ox\sog{\ak{U}+\ak{D}}=\hrt a\uk\ox\sog{{\setlength\arraycolsep{5pt}\begin{array}{cc}%
\co{gt/\h} & \si{gt/\h}\\ %
-\si{gt/\h} & \co{gt/\h}
\end{array}}}_a\sog{\begin{array}{c}%
1\\%
1
\end{array}}_a+\\
&&+\hrt b\dk\ox\sog{{\setlength\arraycolsep{5pt}\begin{array}{cc}%
\co{gt/\h} & -\si{gt/\h}\\%
\si{gt/\h} & \co{gt/\h}
\end{array}}}_a\sog{\begin{array}{c}%
1\\%
1
\end{array}}_a=\\
&&=\hrt
a\uk\ox\sog{\sog{\co{gt/\h}+\si{gt/\h}}\ak{U}+\sog{\co{gt/\h}-\si{gt/\h}}\ak{D}}+\\
&&+\hrt
b\dk\ox\sog{\sog{\co{gt/\h}-\si{gt/\h}}\ak{U}+\sog{\co{gt/\h}+\si{gt/\h}}\ak{D}}.\\
&&\ket{\Psi_{sa}^i(t_1=\frac{\pi\h}{4g})}=a\uk\ox\ak{U}+b\dk\ox\ak{D}.
\end{eqnarray*}}
Derivation of (\ref{eq:a-e}): {\setlength\arraycolsep{0pt}
\begin{eqnarray*}
&&H_{ae}=-\sig_z^{(a)}\ox\sum_{k=1}^Ng_k\sig_{z,k}^{(e)}\prod_{j\neq
k}\ox\mathrm{1}_j.\\
&&\ket{\Psi^i(t_1)}=\sog{a\uk\ox\ak{U}+b\dk\ox\ak{D}}
\prod_{k=1}^N\ox\sog{\al_k\ak{u}_k+\bt_k\ak{d}_k}.\\
&&\ket{\Psi^i(t_1\leq t)}=\left(
a\e{i\sog{\sum_{k=1}^Ng_k\sig_{z,k}^{(e)}\prod_{j\neq
k}\ox\mathrm{1}_j}(t-t_1)/\h}\uk\ox\ak{U}\right.+\\
&&\left.+b\e{-i\sog{\sum_{k=1}^Ng_k\sig_{z,k}^{(e)}\prod_{j\neq
k}\ox\mathrm{1}_j}(t-t_1)/\h}\dk\ox\ak{D}\right)
\prod_{k=1}^N\ox\sog{\al_k\ak{u}_k+\bt_k\ak{d}_k}=\\
&&=a\uk\ox\ak{U}\prod_{k=1}^N\ox\sog{\al_k\e{ig_k(t-t_1)/\h}
\ak{u}_k+\bt_k\e{-ig_k(t-t_1)/\h}\ak{d}_k}+\\
&&+b\dk\ox\ak{D}\prod_{k=1}^N\ox\sog{\al_k\e{-ig_k(t-t_1)/\h}
\ak{u}_k+\bt_k\e{ig_k(t-t_1)/\h}\ak{d}_k}.
\end{eqnarray*}}

\end{document}